\documentclass[preprintnumbers,prl,amsmath,amssymb,superscriptaddress, twocolumn,a4paper,reprint]{revtex4-1}
\usepackage{graphicx}
\usepackage{endnotes}
\usepackage{bm}
\usepackage{epsfig}
\usepackage{dcolumn}
\newcolumntype{d}{D{.}{.}{3.3}}
\newcolumntype{s}{D{.}{.}{2.1}}

\newcommand{\CaSn}{Ca$_3$Ir$_4$Sn$_{13}$}
\newcommand{\SrSn}{Sr$_3$Ir$_4$Sn$_{13}$}
\newcommand{\CaSrSn}{(Ca$_x$Sr$_{1-x}$)$_3$Ir$_4$Sn$_{13}$}

\begin{document}


\title{Pressure-induced and Composition-induced Structural Quantum Phase Transition in the Cubic Superconductor (Sr/Ca)$_3$Ir$_4$Sn$_{13}$}

\author{Lina~E.~Klintberg}
\author{Swee~K.~Goh}
\email{skg27@cam.ac.uk}
\affiliation{Cavendish Laboratory, University of Cambridge, J.J. Thomson Avenue, Cambridge CB3 0HE, United Kingdom}

\author{Patricia~L.~Alireza}
\affiliation{Cavendish Laboratory, University of Cambridge, J.J. Thomson Avenue, Cambridge CB3 0HE, United Kingdom}

\author{Paul J. Saines}
\affiliation{Department of Materials Science and Metallurgy, University of Cambridge, Pembroke Street, Cambridge CB2 3QZ, United Kingdom}

\author{David A. Tompsett}
\altaffiliation{Current address: Department of Chemistry, University of Bath, Bath BA2 7AY, United Kingdom}
\affiliation{Cavendish Laboratory, University of Cambridge, J.J. Thomson Avenue, Cambridge CB3 0HE, United Kingdom}

\author{Peter W. Logg}
\affiliation{Cavendish Laboratory, University of Cambridge, J.J. Thomson Avenue, Cambridge CB3 0HE, United Kingdom}

\author{Jinhu~Yang}
\author{Bin~Chen}
\affiliation{Department of Physics, Graduate School of Science, Hangzhou Normal University, Hangzhou 310036, China}
\affiliation{Department of Chemistry, Graduate School of Science, Kyoto University, Kyoto 606-8502, Japan}

\author{Kazuyoshi~Yoshimura}
\affiliation{Department of Chemistry, Graduate School of Science, Kyoto University, Kyoto 606-8502, Japan}

\author{F.~Malte~Grosche}
\affiliation{Cavendish Laboratory, University of Cambridge, J.J. Thomson Avenue, Cambridge CB3 0HE, United Kingdom}

\date{October 08, 2012}


\begin{abstract}

  We show that the quasi-skutterudite superconductor
  Sr$_3$Ir$_4$Sn$_{13}$ undergoes a structural transition from a
  simple cubic parent structure, the $I$-phase, to a superlattice
  variant, the $I'$-phase, which has a lattice parameter twice that of
  the high temperature phase. We argue that the superlattice
  distortion is associated with a charge density wave transition of
  the conduction electron system and demonstrate that the superlattice
  transition temperature $T^*$ can be suppressed to zero by combining
  chemical and physical pressure. This enables the first comprehensive
  investigation of a superlattice quantum phase transition and its
  interplay with superconductivity in a cubic charge
  density wave system.

\end{abstract}


\maketitle

Structural self-organisation is a central theme in condensed matter
physics. Often, the symmetry of a given parent structure is lowered by
subtle structural variations which decrease the electronic degeneracy
and thereby the total energy. Examples include Jahn-Teller
\cite{Bersuker06} and Peierls distortions \cite{Anderson73} and, more
generally, modulated lattice distortions, or superlattices. A very
diverse family of materials can be explored within the general
R$_3$T$_4$X$_{13}$ stoichiometry, where R is an earth
alkaline or rare-earth element, T is a transition metal and X is a
group-IV element. Among these are the superconducting and magnetic
stannides \cite{remeika80,espinosa80}
, Ce-based Kondo lattice systems \cite{sato93}, and
thermoelectrics \cite{strydom07}. Many members of this family adopt a
variant structure, the $I'$-phase, derived from the simple cubic
parent structure ($I$-phase, $Pm\bar3n$). Empirically, compounds with
divalent or tetravalent cations R occur in the $I$-phase, whereas
compounds with trivalent cations form in the $I'$-phase. Detailed
diffraction studies \cite{hodeau82,miraglia86,bordet91} disagree about
the precise $I'$-phase structure, but consistently interpret it as a deformation of the X$_{12}$ cages, which can be viewed
as a superlattice distortion of the $I$-phase with twice the original
lattice constant.

A recent reexamination \cite{yang10} of Ca$_3$Ir$_4$Sn$_{13}$, which
superconducts below $T_c = 7~{\rm K}$, revealed distinct anomalies in
the electrical resistivity and magnetic susceptibility at
$T^*\simeq33~\rm K$. In the isoelectronic sister-compound
Sr$_3$Ir$_4$Sn$_{13}$, the equivalent anomaly occurs at
$T^*\simeq147~\rm K$ and superconductivity sets in at $T_c = 5~{\rm
  K}$. Intermediate compositions (Ca$_x$Sr$_{1-x}$)$_3$Ir$_4$Sn$_{13}$
form readily. This invites studies which exploit the negative chemical
pressure from partial substitution of Ca by Sr, together with positive
physical pressure from a hydrostatic pressure cell.  In this Letter,
we report a detailed investigation of the nature of the transition at
$T^*$ by x-ray diffraction (XRD), and we examine the dependence of
$T^*$ and of the superconducting and normal state properties on
physical and chemical pressure. We find that (i) the $T^*$ anomaly is
produced by a second order superlattice transition into the
$I'$-phase, (ii) $T^*$ is suppressed with increasing pressure and
extrapolates to zero temperature at $p_c \simeq 18 ~\rm {kbar}$ in
Ca$_3$Ir$_4$Sn$_{13}$, 
and (iii) $T_c$ peaks near $p_c$, and the electrical 
resistivity adopts a linear temperature dependence over a wide
range of temperature and magnetic field close to this critical pressure.

The \CaSrSn\ single crystals were grown by a flux method
\cite{espinosa80}.  Four-wire AC resistivity measurements were performed
in a piston-cylinder cell, using a Physical
Property Measurement System (Quantum Design) to control temperature,
$T$. Two Moissanite anvil cells were prepared for AC susceptibility
measurements with a conventional mutual inductance method, in which a
10-turn microcoil \cite{Goh08, Alireza03, Klintberg10} is placed
inside the gasket hole (thickness: 150~$\mu$m, diameter: 400~$\mu$m)
as the pickup coil. Glycerin was used as the pressure medium for the
piston-cylinder cell and for one of the anvil cells, and 4:1
methanol-ethanol mixture was used for the other anvil cell. Ruby
fluorescence spectroscopy and the $T_c$ of lead were used to determine
the pressure in the anvil and piston-cylinder cell
respectively. Single crystal XRD measurements were performed using an
Oxford Diffraction Gemini E Ultra utilising MoK$\alpha$ radiation at
100 to 295~K. The structures were solved using direct methods and
refined using Shelx-97 \cite{Sheldrick} via the WinGX interface
\cite{Farrugia}.  The electronic structure was calculated using the Local Density Approximation (LDA) and
the Generalized Gradient Approximation (GGA) \cite{Perdew96} with Wien2k
\cite{Wien}. 
$Rk_{max}\!=\!7$ and 40,000 $k$-points were used in a
non-spin polarized calculation. The position of the 24$k$ Sn site (0,
y, z), the only free internal coordinate, was optimised numerically,
resulting in (0, $y\!=\!0.3045$, $z\!=\!0.1516$) and (0,
$y\!=\!0.3027$, $z\!=\!0.1522$) for Ca- and \SrSn\ respectively.

\begin{figure}[t]
\includegraphics[width=\columnwidth]{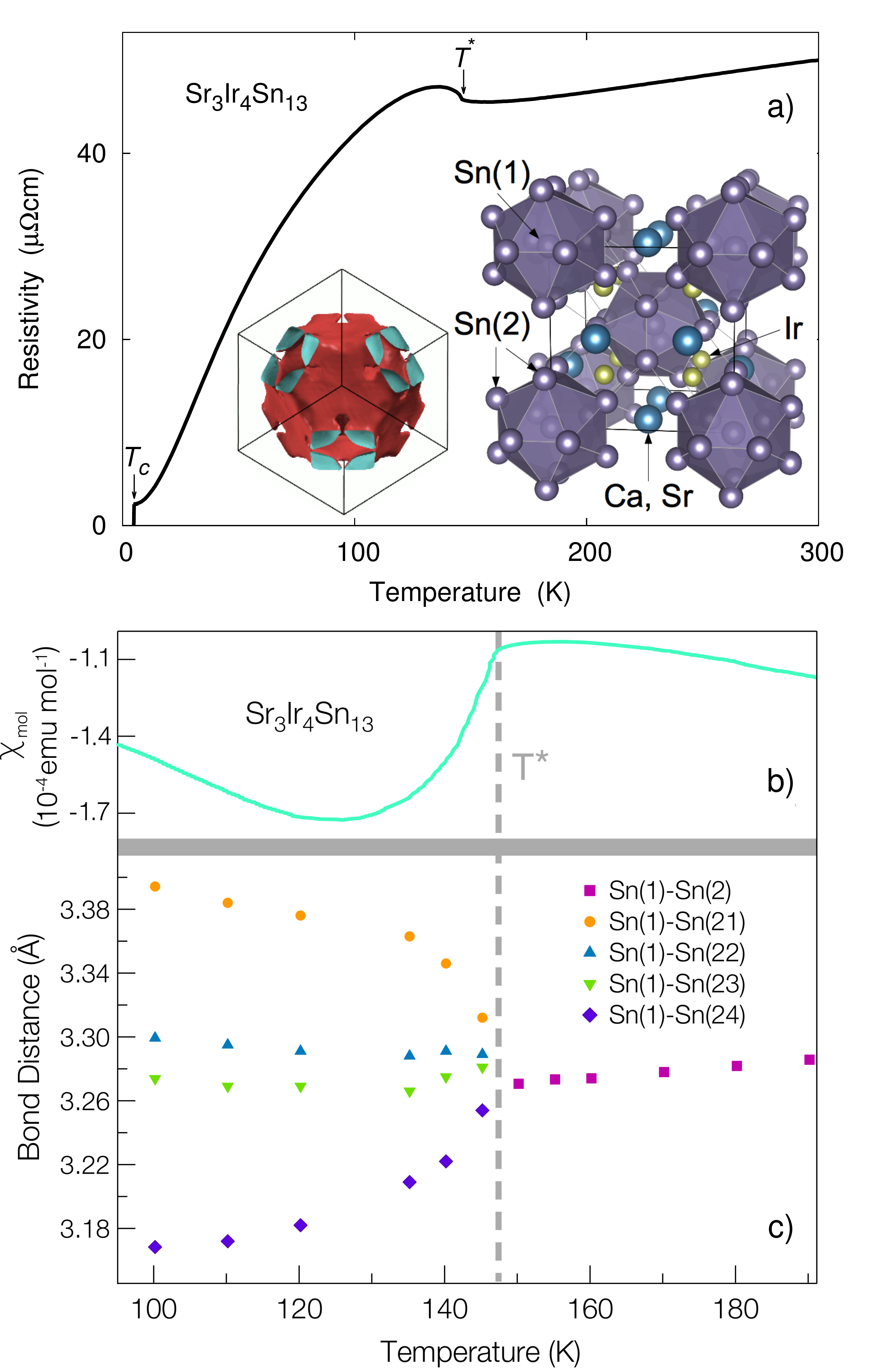}
		
\caption{\label{SISTransition} Key properties of Sr$_3$Ir$_4$Sn$_{13}$
  at ambient pressure. The main panels show the temperature dependence
  of (a) the electrical resistivity, (b) the magnetic susceptibility,
  and (c) interatomic bond-lengths. The right inset to panel (a)
  illustrates the crystal structure of Sr$_3$Ir$_4$Sn$_{13}$,
  featuring Sn(2) 12-cages (icosahedra), which each contain a central
  Sn(1) atom. The bond-length between the central Sn atom and the cage
  atoms the $I$-phase, splits into four distinct lengths in the
  $I'$-phase below $T^*\simeq147~\rm K$. The symbol sizes represent
  the measurement accuracy. The left inset to panel (a) shows one of
  the larger Fermi surface sheets (band 329) computed in the $I$-phase of
  Sr$_3$Ir$_4$Sn$_{13}$.}
\end{figure}

Room temperature single crystal XRD confirms that
Sr$_3$Ir$_4$Sn$_{13}$ forms in the $I$-phase structure
(\textit{Pm\=3n}, lattice parameter $a = 9.7968(3) ~{\rm \AA}$ at
$295~{\rm K}$, inset of Fig.~\ref{SISTransition}a). On cooling,
distinct signatures of a phase transition at $T^* = 147~\rm K$ are
observed in the temperature dependence of the electrical resistivity
$\rho$ and of the magnetic susceptibility $\chi$ of
Sr$_3$Ir$_4$Sn$_{13}$, in addition to a
superconducting transition at $T_c =5~\rm K$ (Fig.~\ref{SISTransition}).  Below $T^*$ our single
crystal XRD data are inconsistent with the $I$-phase. They must
instead be indexed  \cite{SupMat} to the $I'$-phase, a body(I)-centered cubic
structure with a lattice parameter twice that of the
high temperature phase  (\textit{I\=43d}, $a=19.5947(3)~{\rm \AA}$ at
$100~{\rm K}$, $R$1, $wR$2 and $\chi^2$ of 4.4~\%, 13.9~\%
and 1.1~\%, respectively).  In the $I$-phase, the bond distances in
the Sn(1)Sn(2)$_{12}$ icosahedra and in the IrSn(2)$_6$ trigonal
prisms which connect them are identical. In the $I'$-phase, however, a
distortion of the Sn(2) icosahedra gives rise to four groups, Sn(1)-Sn(21-24), each group composed of three bonds with identical bond distances (see Fig.
\ref{SISTransition}c). This distortion occurs in concert with tilting
of three-quarters of the trigonal prisms while those whose axis are
along the (111) direction remain untilted.
Because of the periodic nature of this distortion, it leads to the formation of a superlattice.

\begin{table}[tb]
  \caption{Interplay between lattice and electronic structure in Sr$_3$Ir$_4$Sn$_{13}$, investigated by \textit{ab initio} simulation using density functional theory (DFT) with local (LDA) and gradient-corrected (GGA) semi-local functionals. Lattice parameters $a$ 
    and electronic density of states at the Fermi energy $g(E_F)$ have been computed after full structural relaxation in both the $I$ and $I'$ phase. $\Delta E$  is the difference in total energy between the two phases. The lower DOS and total energy obtained for the $I'$-phase agree with the experimental observation that Sr$_3$Ir$_4$Sn$_{13}$ adopts the $I'$ structure at low temperature, and the small magnitude of $\Delta E$ is consistent with the low transition temperature $T^*$. }
\label{table1}
\begin{ruledtabular}
\begin{tabular}{c d s s d s s}	
& \multicolumn{3}{c}{GGA calculation} &  \multicolumn{3}{c}{LDA calculation}  
\\

 & \multicolumn{1}{c}{$a$}  & \multicolumn{1}{c}{$g(E_F)$} & \multicolumn{1}{c}{$\Delta E$} & \multicolumn{1}{c}{$a$}  & \multicolumn{1}{c}{$g(E_F)$} & \multicolumn{1}{c}{$\Delta E$}
\\
 
& \multicolumn{1}{c}{${\rm \AA}$} & \multicolumn{1}{c}{$({\rm f.u.~eV})^{-1}$} & \multicolumn{1}{c}{${\rm meV / f.u.}$} & \multicolumn{1}{c}{${\rm \AA}$} & \multicolumn{1}{c}{$({\rm f.u.~eV})^{-1}$} & \multicolumn{1}{c}{${\rm meV / f.u.}$}
\\

\colrule 
$I$ & 9.937 & 12.5 &  & 9.698 & 12.5 &   
\\

$I'$ & 2 \cdot 9.941 & 9.3  & -7.1 & 2 \cdot 9.700 & 8.5 & -3.4 
\end{tabular}
\end{ruledtabular}
\end{table}

Electronic structure calculations give further insight into the nature
of this superlattice transition. Our calculations (Table
~\ref{table1}) suggest that in Sr$_3$Ir$_4$Sn$_{13}$ the $I'$-phase
superlattice has a slight energy advantage with respect to the
$I$-phase parent structure.  Moreover, the period-doubling associated
with the superlattice causes the large Fermi surface sheets
computed for the $I$-phase (e.g. left inset in
Fig.~\ref{SISTransition}a) to reconstruct,  which gaps out significant sections of the Fermi
surface. The calculated reduction in the
electronic density of states at the Fermi energy amounts to nearly $30 \%$, or $\simeq 3.5~{\rm
  eV^{-1}}$ per formula unit, which translates to a change in the Pauli
susceptibility of $\Delta \chi = \mu_0 \mu_B^2 g(E_F) \simeq 1.1\cdot
10^{-4} ~{\rm emu/mol}$. This value is consistent with the
observed reduction of the measured susceptibility at $T^*$
(Fig.~\ref{SISTransition}b). The distinct increase of  $\rho(T)$ below
$T^*$ may also be attributed at least in part to the reduction in $g(E_F)$, along
with other factors, such as changes in the effective carrier mass
and the scattering rate. 

The significant reduction of $g(E_F)$ on entering the $I'$-phase
indicates that electronic states near the Fermi energy play an
important role in forming the superlattice, which suggests a charge
density wave (CDW) instability of the conduction electron system. In
the $I$-phase, six bands cross the Fermi level. Each sheet of the
Fermi surface is three dimensional, but some sheets exhibit low
curvature regions with the necessary characteristics for nesting: the
flat sections of band 329 (inset of Fig.~\ref{SISTransition}a) are
strong candidates. The contribution of this band to the real part of
the wavevector-dependent charge susceptibility $\chi({\bf q})$, or
Lindhard function, peaks at ${\bf q} =(1/2, 1/2, 1/2)$, with a 21\%
enhancement above that at the Brillouin zone centre \cite{SupMat}.
This confirms the visual impression that sections of the Fermi surface
nest along the body diagonal and is consistent with the experimentally
observed doubling of the lattice parameter in all directions. As in
most CDW systems \cite{Johannes08}, the response of the conduction
electrons in isolation is not singular, but strong electron-phonon
interactions can induce an instability at the wavevector selected by
the peak in the Lindhard function. Our calculations suggest that the
low temperature $I'$-phase is stabilised by the interplay between a
periodic lattice distortion and a CDW in the conduction electron system.

\begin{figure}[t]
 \includegraphics[width=\columnwidth]{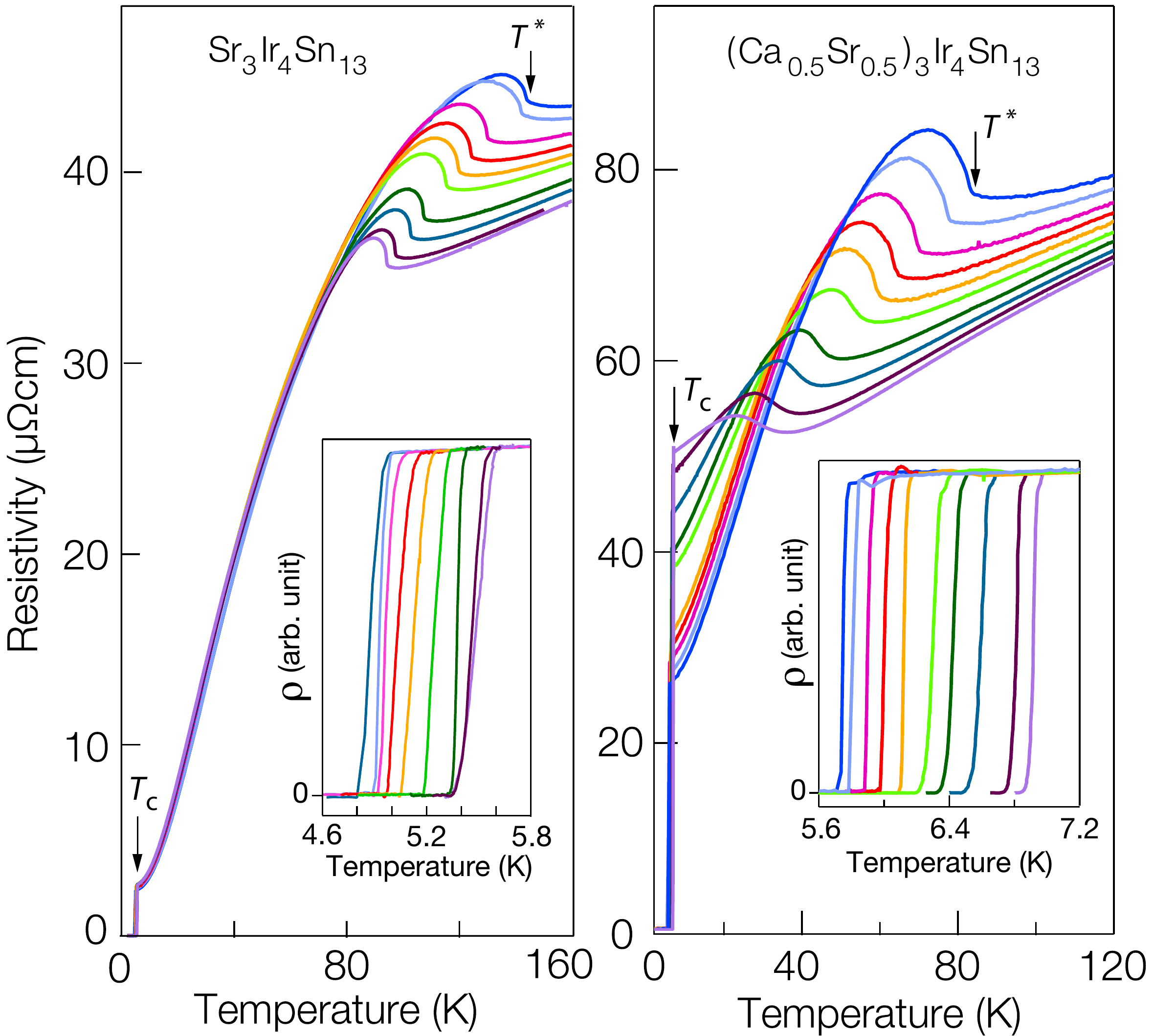}
             				
 \caption{High pressure resistivity data in Sr$_3$Ir$_4$Sn$_{13}$
   (left) and in mid-way Ca-substituted
   (Sr$_{0.5}$Ca$_{0.5}$)$_3$Ir$_4$Sn$_{13}$, showing both the
   superlattice and superconducting transitions. The pressures are
   given by 1.1, 3.8, 8.7, 11.7, 13.7, 17.3, 19.3, 21.8, 23.5, 25.7
   kbar in sequential order with the highest $T^*$ at the lowest
   pressure. The insets show expanded views of the superconducting transitions.}
\label{SISPressure}
\end{figure}

\begin{figure}[t]
 \includegraphics[width=0.98\columnwidth]{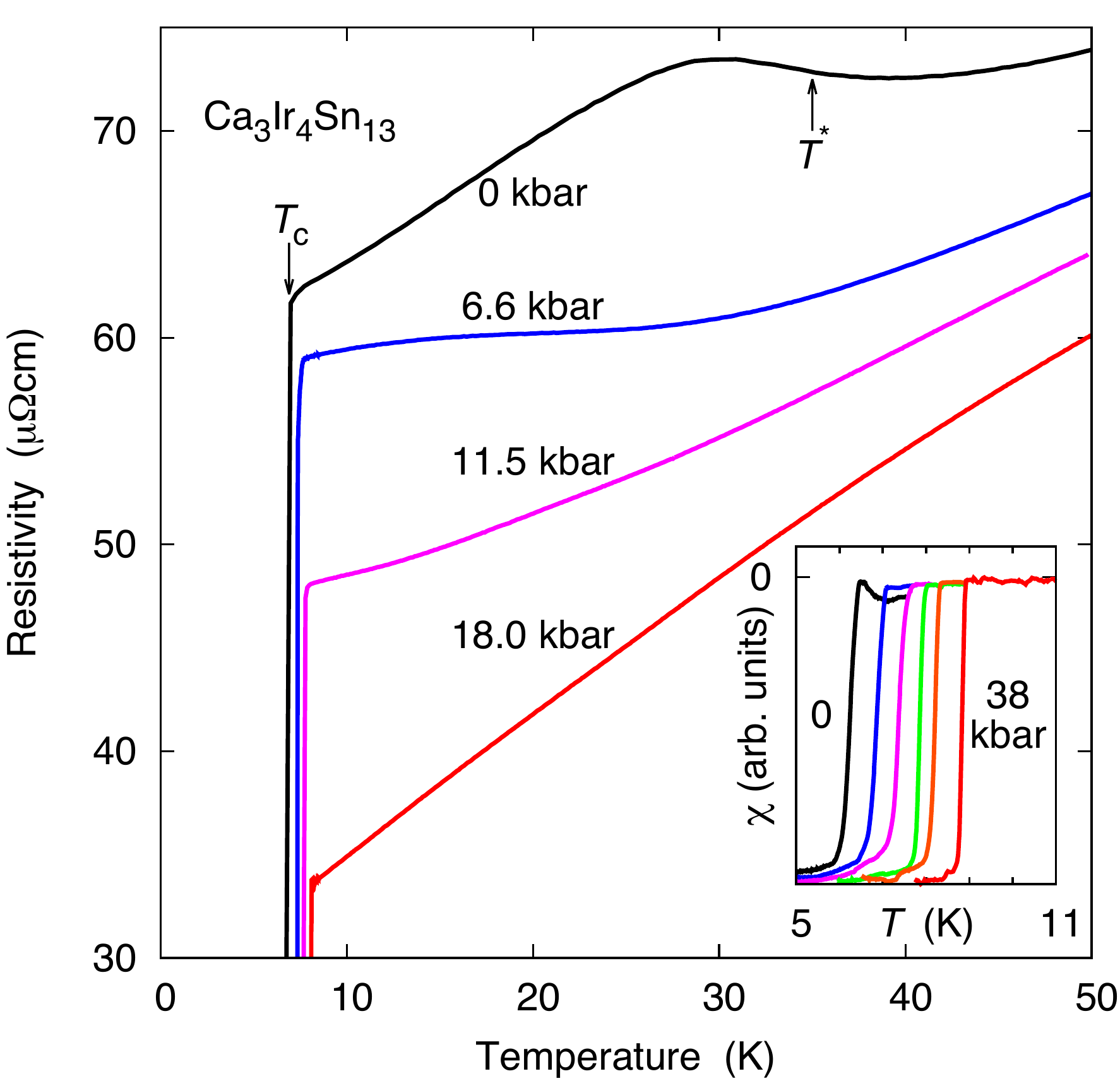}             				
 \caption{ (Main panel) Temperature dependence of the electrical
   resistivity in \CaSn\ at different applied
   hydrostatic pressures. $T^*$ is defined as the minimum of
   $d\rho/dT$, which broadens on approaching the critical
   point
   . Near the critical pressure $p_c\simeq18$ kbar, where $T^*$
   extrapolates to 0 K, $\rho(T)$ assumes a linear temperature
   dependence over a large temperature range. (Inset) Superconducting
   transitions in \CaSn observed in the magnetic susceptibility at,
   from the left, 0, 7, 57, 48, 28 and 38 kbar. $T_c$ peaks near 8.9 K at about 40 kbar.}
\label{CISPressure}
\end{figure}

\begin{figure}[t]
\includegraphics[width=\columnwidth]{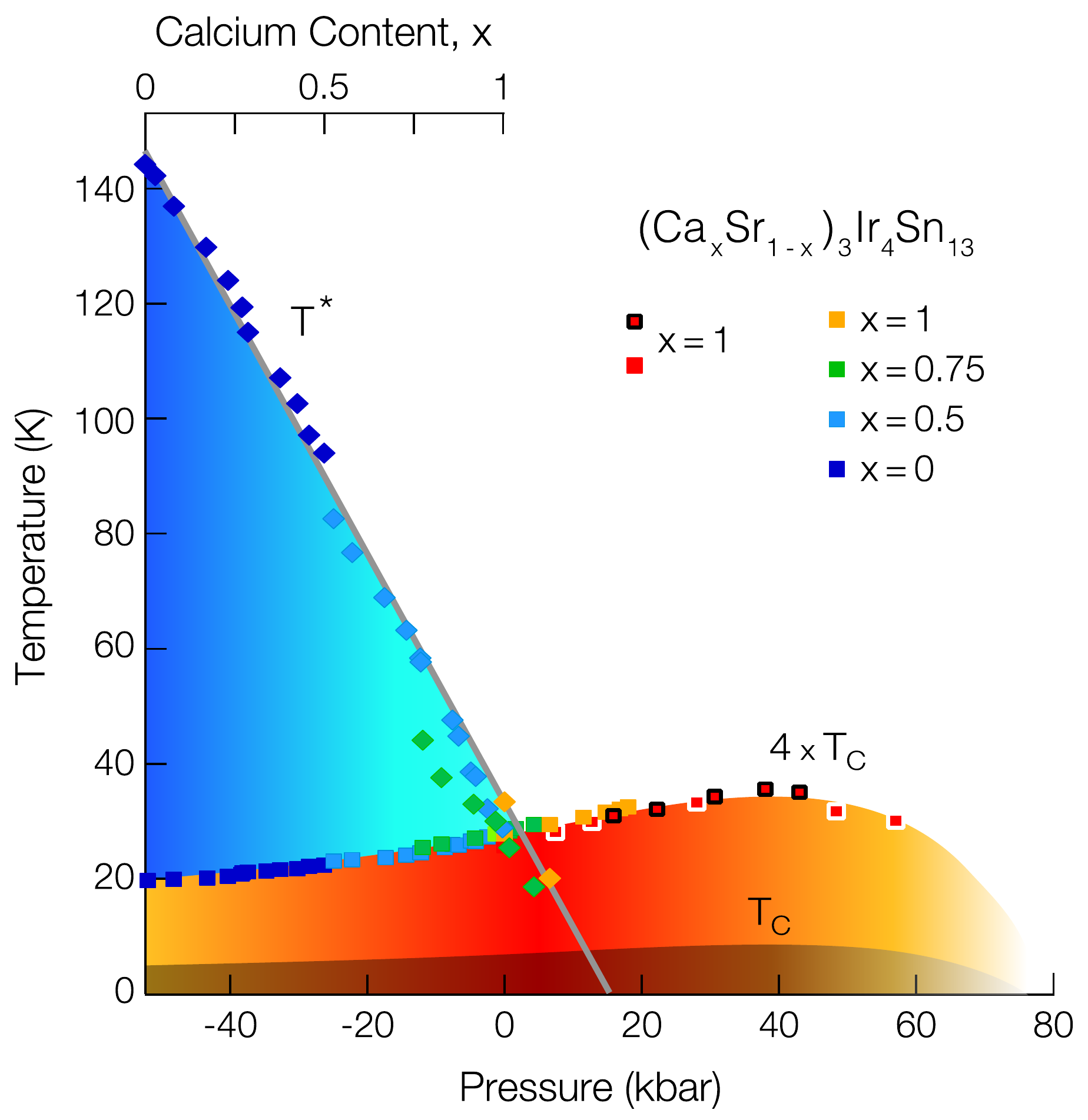}        				
\caption{Universal phase diagram for the \CaSrSn ~ system, constructed
  by placing $x=0$, 0.50 and 0.75 at $-52$, $-26$ and $-13$~kbar,
  respectively (c.f. top axis). The superlattice transition
  temperature, $T^*$, extrapolates to zero temperature at roughly
  18~kbar. The red ($x\!=\!1$) symbols denote values derived from AC
  susceptibility measurements -- the white (black) borders represent a
  pressure medium of glycerin (4:1 methanol:ethanol). The remaining
  data are obtained from the analysis of resistivity measurements. The
  superconducting transition temperature, $T_c$, is multiplied by 4
  for clarity.  }
\label{PhaseDia}
\end{figure}

High-pressure measurements of the resistivity, $\rho(T)$, in
Sr$_3$Ir$_4$Sn$_{13}$ show that $T^*$ decreases rapidly with
increasing pressure, $p$, whereas $T_c$ rises slowly
(Fig.~\ref{SISPressure}a). This pattern carries over to partially
Ca-substituted samples such as
(Sr$_{0.5}$Ca$_{0.5}$)$_3$Ir$_4$Sn$_{13}$ (Fig.~\ref{SISPressure}b),
which due to its smaller unit cell volume can be regarded as a high
pressure analogue of Sr$_3$Ir$_4$Sn$_{13}$. In the end member of this
series, Ca$_3$Ir$_4$Sn$_{13}$ (Fig.~\ref{CISPressure}), the
resistivity anomaly associated with $T^*$ broadens and $T^*$
decreases further with increasing pressure, extrapolating to 0 K at a
critical pressure $p_c \simeq 18 ~{\rm kbar}$ (see also
Fig.~\ref{PhaseDia}). This constitutes a structural quantum phase
transition or, if the transition remains continuous, a structural
quantum critical point.  Near $p_c$, $\rho(T)$ is linear over a wide
temperature range, from $T_c$ up to 50 K. Anvil cell AC susceptibility
measurements extend the available pressure range and show that $T_c$
peaks near $8.9~\rm K$ at $\simeq 40~\rm {kbar}$ (inset of
Fig.~\ref{CISPressure}).

Analysis of the pressure dependence of $T^*$ and $T_c$ at intermediate
substitution values $x$ suggests that the effect of
fully substituting Ca with Sr places \SrSn\ at roughly $-52~\rm{kbar}$
relative to the physical pressure scale of
Ca$_3$Ir$_4$Sn$_{13}$. Applying Vegard's law  
to intermediate composition values
 allows us to construct a universal temperature-pressure phase diagram
 for the (Ca$_x$Sr$_{1-x}$)$_3$Ir$_4$Sn$_{13}$ series
 (Fig.~\ref{PhaseDia}). The observation of a superconducting dome,
 which peaks close to $p_c$, and the anomalous quasi-linear $\rho(T)$ are
 often associated with quantum critical phenomena on the threshold of
 magnetism \cite{mathur98}. There is, however, no evidence that the
 low temperature order observed in (Sr/Ca)$_3$Ir$_4$Sn$_{13}$ is
 magnetic. On the contrary, the fact that no magnetic order has been
 reported in any other $I'$-phase material without rare-earth
 constituents, the low absolute magnetic susceptibility values, the
 absence of clear anisotropy in the susceptibility at $T^*$, our spin
 polarised band structure calculations within density functional
 theory for various hypothetical magnetic states, in which the
 magnetic moments collapse to zero in all cases, and the strong
 indications that the superconducting state is conventional and fully
 gapped \cite{Kase11,zhuo12} all point towards a non-magnetic
 transition at $T^*$. Because $g(E_F)$ is strongly reduced in the
 $I'$-phase (Table \ref{table1}), conventional BCS theory suggests
 that the associated $T_c$ is low compared to the value that could be
 achieved in the $I$-phase. Conversely, when $T^*$ is suppressed with
 hydrostatic pressure, and the $I$-phase survives to low temperatures,
 $g(E_F)$ increases. This would be expected to raise $T_c$ on
 approaching $p_c$, as is indeed observed.

To understand the initial further increase of $T_c$ for $p>p_c$, the decrease of
$T_c$ at much higher pressures and the linear temperature dependence of $\rho(T)$ near $p_c$
(Fig.~\ref{CISPressure}) we need to look beyond $g(E_F)$ and examine
the evolution of the phonon spectrum.
If the superlattice
transition remains second order or only weakly first order, then the
associated optical phonon mode should soften at $T^*$, which itself
approaches zero at $p_c$. This generates a low-lying, weakly
dispersive phonon
branch at low temperature. We would expect its contribution to the
electrical resistivity to become linear in temperature, in analogy
with the behaviour of simple metals above the Debye temperature, once
the thermal energy $k_B T$ exceeds the maximum phonon energy of this
branch. A more detailed argument along the lines presented for Einstein
solids \cite{cooper74,
  lortz06}, notes that the
phonon contribution to the electrical resistivity can be written as 
\begin{equation}
\Delta\rho_{ph}(T) \propto \sum_{\bf q} \alpha_{(tr) \bf q}^2 T \left(\partial n_{\bf
  q}/\partial T\right)_{\omega_{\bf q}} \quad .
\label{Tlinear}
\end{equation}
The sum is taken over all phonon wavevectors $\bf q$ within a
suitable cut-off, and $\alpha_{(tr) \bf q}^2$ is a ${\bf
  q}$-dependent Fermi surface average of the 
electron-phonon interaction, which is proportional to the density of
states at the Fermi energy, has been weighted appropriately for
transport calculations, and depends weakly on ${\bf q}$ away
from $q=0$. Moreover,
$\omega_{\bf q}$ denotes the ${\bf
  q}$-dependent phonon frequency, and
$n_{\bf q} = \left[\exp(\hbar \omega_{\bf q}/(k_B T)) -1 \right]^{-1}$ is
  the Bose occupation number. When $k_B T$ exceeds the highest
  $\hbar \omega_{\bf q}$, this expression reverts to $T \sum
  \omega_{\bf q}^{-1}$, giving a $T-$linear resistivity, which is
  strongly enhanced for low $\omega_{\bf q}$. Because the  electron-phonon coupling
constant $\lambda$ can be expressed as $\lambda=2 \sum_{\bf q} \alpha_{\bf q}^2
  \omega_{\bf q}^{-1}$ (see, e.g. \cite{mcmillan68}), where
  $\alpha_{\bf q} \simeq \alpha_{(tr) \bf q}$ for 
  large $q$, it is seen to be directly connected to the slope of the $T-$linear resistivity.

The softening of parts of the phonon spectrum can also help explain the
shape of the superconducting dome (see, e.g. \cite{calandra11}). In the Eliashberg-McMillan
treatment \cite{mcmillan68} $T_c$ is exponentially sensitive to $\lambda$, which in
turn is proportional to $g(E_F)$ and to $\sum
  \omega_{\bf q}^{-1}$.
As $T^*$ is reduced, the removal of Fermi surface reconstruction
raises $g(E_F)$, while the
softening of parts of the phonon spectrum increases $\sum
  \omega_{\bf q}^{-1}$. Both effects tend to increase $\lambda$, which, if all other parameters are constant, raises
$T_c$. 
Beyond these considerations, the precise effect of a soft mode on $T_c$ has
to be considered in order to understand the shape of the superconducting dome
for pressures above $p_c$. Whereas $T_c$ depends exponentially on
$\lambda$, its scale prefactor is set by a weighted average of $\omega_{\bf q}$. The combined effect of scale and exponent is not
necessarily optimised at the point where an optical mode is soft, as
this minimises the scale prefactor. We would, instead, expect the peak
to occur slightly beyond the critical point in the
phase diagram, as found experimentally (Fig.~\ref{PhaseDia}).
Eventually, the optical mode associated with the superlattice
transition stiffens sufficiently so that the detrimental effect this
has on $\lambda$ begins to reduce $T_c$. Detailed
theoretical calculations based on more complete information about the
phonon spectrum will be required to arrive at a quantitative
understanding of the observed phase diagram.

Investigations into the effect of lattice instabilities on
superconductivity have a long history, starting with the structurally
related A15 compounds \cite{testardi75,tanaka10}. Examples of enhanced
superconductivity near lattice instabilities include
Lu$_5$Ir$_4$Si$_{10}$ \cite{Shelton86}, dichalcogenides such as
TiSe$_2$ \cite{Morosan06,Kusmartseva09} and TaS$_2$ \cite{Sipos08},
intercalated graphite CaC$_6$ \cite{gauzzi07}, 1- or 2-D organic
compounds \cite{Lubczynski96,Wosnitza01}, and a number of elements at
very high pressure \cite{degtyareva07}. There is an emerging view that
in most cases the evolution of the phonon spectrum with tuning
parameters like pressure crucially influences $T_c$, both by boosting
it when the lattice instability is suppressed, and by reducing it
again when pressure is increased further. Whereas most of the
aforementioned materials are low-dimensional, very few clear cases of
CDW order in 3D materials exist, notably cubic CuV$_2$S$_4$
\cite{fleming81}, which is not superconducting, and orthorhombic
$\alpha-$U \cite{Lander94}, for which a recent detailed examination of
the phonon spectrum \cite{raymond11} suggests similarities to the
mechanisms outlined above. (Sr/Ca)$_3$Ir$_4$Sn$_{13}$ allows us to
investigate the interaction between a structural instability and
superconductivity in a cubic material. Our study clarifies the nature
of the previously unidentified phase transition in
(Sr/Ca)$_3$Ir$_4$Sn$_{13}$, it demonstrates that this transition can
be tuned to zero temperature, suggesting a structural quantum critical
point, and it attributes the linear $\rho(T)$ and the dome structure
of $T_c$ to the associated softening of optical phonon modes. The
prospect of fine-tuning phonon frequencies by controlling a structural
transition temperature motivates further studies in this material
class as well as in more complex systems, in which spin and charge
density wave transitions may be correlated.

\textbf{Acknowledgements} We particularly thank G. G. Lonzarich,
C. Pickard, and D. Khmelnitskii for helpful discussions. This work was supported by the EPSRC
UK, Trinity College (Cambridge), Grant-in-Aid for Scientific Research
from the JSPS (22350029), and the Global COE Program ''International Center
for Integrated Research and Advanced Education in Materials Science''
at Kyoto University. SKG acknowledges the Great
Britain Sasakawa Foundation for travel support and Kyoto University for hospitality.


\end{document}